# Introduction to Multilevel Modeling Techniques

**Amira. El-Deosokey**

Assistant Professor, Higher Future Institute for specialized technological Studies, Egypt.

## Abstract

In this paper, I outline several conceptual and methodological issues related to modeling individual and group processes embedded in clustered/hierarchical data structures. We position multilevel modeling techniques within a broader set of univariate and multivariate methods commonly used to study different types of data structures. We then illustrate how the choice of analysis method affects how best to examine the data. This overview gives us an idea of our further development of these themes and models in this study.

## Introduction

Over the past several decades, concerns in various fields with conceptual and methodological issues in conducting research with hierarchical (or nested) data have led to the development of multilevel modeling techniques. Research on organizations such as universities or product and service firms presents opportunities to study phenomena in hierarchical settings. Individuals (Level 1) may work within specific formally defined departments (Level 2), which may be found within larger organizations (Level 3), which, in turn, may be located within specific states, regions, or nations. These individuals interact with their social contexts in a variety of ways. Individuals bring certain skills and attitudes to the workplace; they are clustered in departments or work units having certain characteristics, and they are also clustered within organizations having particular characteristics. Because of the presence of these successive groupings in hierarchical data, individuals within particular organizations may share certain properties including socialization patterns, traditions, attitudes, and work goals. Similarly, properties of groups (e.g., leadership patterns, improvement in productivity) may also be influenced by the people in them. Hierarchical data also result from the specific research design and the nature of the data collected. In survey research, for example, individuals are often selected to participate in a study from some type of stratified random sampling design (e.g., individuals may be chosen from certain neighborhoods in particular cities and geographical areas). Longitudinal data collection also presents another research situation where a series of measurements is nested within the individuals who participate in the study.

In contrast, *disaggregation* refers to moving a variable conceptualized at a higher level to a lower level. For example, in a different analysis we may have productivity measured at the organizational level but also have items that express individual employee attitudes and motivation. In this case, we intend to analyze the data at the individual level to see whether employee attitudes influence productivity. If we assign to all individuals the same value on the organizational productivity variable (and possibly other organizational variables such as size), we attribute properties of the organization to individuals. This can also confound the analysis.

**MACRO LEVEL**
Context

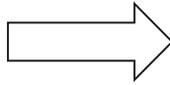

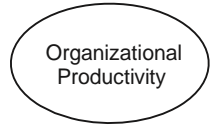
Organizational Productivity

Composition
Structure
Resources

*What contextual, structural, compositional, and resource variables affect organizational productivity?*

---

**MACRO LEVEL**

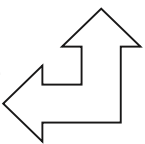
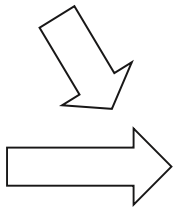
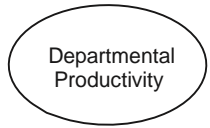
Departmental Productivity

*How do structural characteristics, compositional variables, and teamwork affect departmental productivity?*

---

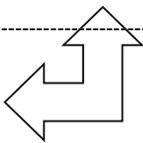
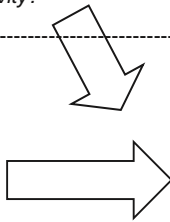
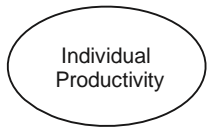
Individual Productivity

Context
Composition

**FIGURE 1.1** Defining variables and relationships in a multilevel conceptual model.

In addition, more recently Muthén (2008) has proposed a variety of hybrid latent variable formulations that include both continuous and categorical latent variables (e.g., latent class analysis, growth mixture analysis, factor mixture analysis). These techniques, which are readily available in the Mplus statistical package (Muthén & Muthén, 1998–2012), expand the manners in which cross-sectional,

**TABLE 1.1** Summary of Quantitative Approaches to the Analysis of Organizational Data

| Analytic Approach | Example Techniques |
| --- | --- |
| *Single-level data structure* | |
| 1. Univariate (one dependent variable) | correlation, analysis of variance (ANOVA), regression, repeated measures ANOVA |
| 2. Multivariate (two or more dependent variables) | canonical correlation, multivariate analysis of variance and discriminant analysis, factor analysis, path analysis, time series analysis, covariance structure models, other types of structural equation models (e.g., latent curve, mixture, latent class) |
| *Multilevel data structure* | |
| 3. Univariate (one dependent variable) | multilevel regression or random coefficients models, variance components models, mixed linear models, time series analysis, growth curve models |
| 4. Multivariate (two or more dependent variables) | multilevel multivariate models, multilevel covariance structure models, other types of multilevel structural equation models (e.g., multilevel latent curve, mixture, latent class) |

longitudinal, and hierarchical data structures may be examined in the coming years. As Muthén noted, the latent variable emphasis of SEM actually provides a general modeling conceptual framework that incorporates **random effects** and univariate outcomes in hierarchical settings as one of several types of models

[e.g., see also Raudenbush & Bryk (2002), Mehta & Neale (2005), and Muthén & Asparouhov (2011) for further discussion].

In Table 1.1, we locate multilevel modeling within a larger methodological framework of quantitative methods of analysis relevant to social and behavioral research. For ease of presentation, we group the methods by data structure (e.g., single-level versus nested or hierarchical) and number of dependent variables (univariate versus multivariate).Within this larger framework, we can identify four general types of analyses involving (1) single or (2) multiple dependent variables and (3) single-level or (4) nested data structures. In general, these types of quantitative techniques, as well as the various hybrid techniques, can all be subsumed under a broader latent variable framework (Muthén, 2002; Muthén & Asparouhov, 2011).

The choice of analytic strategies and model specification is not a trivial one. More complete modeling formulations may suggest inferences based on relationships in the sample data that are not revealed in more simplistic models. At the same time, however, better developed modeling formulations are also more likely to lead to fewer findings of substance than have often been claimed in studies that employ more simplistic analytical methods (Pedhazur & Schmelkin, 1991). We draw a clear distinction between concerns over model specification limited to the exclusion or inclusion of theoretically relevant variables and model specification related to the mathematical explication of the relationships among those variables. While theory should guide both considerations, the former concern deals with the availability of relevant variables in the data set, and the latter deals with the structure of the data set itself and the choice of modeling approach used to exploit theoretically important relationships presumed to exist in the population. Besides the choices we make about methods of analysis, our inferences may also be affected in practice by limitations of our samples (e.g., size and sampling variation, missing data). It is important to acknowledge that these limitations can also lead to potential biases in our results.

**Univariate Analysis**

In the remainder of this chapter, we present several examples to illustrate the potential implications of decisions about methods of analysis and data structures that affect the interpretation of results. Another of our guiding principles is that the responsible researcher should consider approaches that are likely to take full advantage of the features of particular data structures and goals of the overall research when making decisions about analytic methods. We illustrate our point about decisions regarding methods of analysis and fully exploiting features of our data with a series of short examples.

For our first example, let's consider a case where the goal is to examine whether gender is related to student achievement. There are a number of different options in Table 1.1 we could choose to analyze the data in this simple proposed model. One way to think about statistical modeling is in terms of an attempt to account for variation in a dependent variable such as student achievement—variance that is believed to be associated with one or more explanatory variables such as gender and other demographic categories (e.g., socioeconomic status, race/ethnicity) or personal attributes measured as continuous variables (e.g., motivation, previous learning). This is analogous to thinking about how much variance in student achievement ($R^2$) is accounted for by a given set of explanatory variables.

In this case, we might propose that gender accounts for variation in students' test scores. In Table 1.2, we have data compiled on a small random sample of 14 students from a larger study. The data consist of their scores in a reading test, a math test, and a language skills test. The data in the table show that females in the sample scored lower than males on each test. For ease of presentation we set aside the issue of whether there are other variables that should also be included in the model to provide a more thorough examination of whether gender differences in achievement would exist after other known variables related to achievement were controlled.

**TABLE 1.2** Descriptive Statistics for Example

| Variable | Male | | Female | |
| --- | --- | --- | --- | --- |
| | Mean | SD | Mean | SD |
| Reading | 650.88 | 37.01 | 641.50 | 15.80 |
| Math | 679.75 | 36.54 | 637.00 | 15.80 |
| Language | 658.13 | 27.78 | 631.67 | 25.57 |

### *Multiple Regression*

We might choose univariate regression to conduct the analysis. Regression analysis employs cross-sectional data that are obtained as individual-level data through random sampling or individual and group data using cluster sampling. Gender would be the independent variable, and each of the tests would represent a single dependent variable. This strategy would require three separate tests and would get at the issue of how gender affects achievement, but in each case, achievement would be described somewhat differently (i.e., math, reading, language). The ratio of the estimate to its standard error ($\beta$/SE) can be used to provide a *t*-test of statistical significance for each of the hypothesized relationships. Assuming the data in the example are single level, the hypothesis tested is that the population from which males and females were selected has the same means for each of the dependent variables. If we examine the regression coefficients between gender and achievement in Table 1.3, we can see that for this small sample, females scored significantly lower than males in math (unstandardized $\beta = -42.75$) but not in reading ($\beta = -9.38, p > 0.05$) or language ($\beta = -26.46, p > 0.05$). If we were to summarize these data, we would likely conclude that gender affects achievement under some conditions but not others.

### *Analysis of Variance*

Another way to think about statistical modeling is in terms of an attempt to decompose variability in the test score into its component parts. For example, if we used a simple one-way ANOVA to investigate the relationships in Table 1.3, we would be testing the similarity of group means for males and females by partitioning the sum of squares for individuals into a portion describing differences in achievement variability due to groups (i.e., gender) and differences in variability due to individuals. In this case, the *F*-ratio provides an indication of the ratio of between-groups variability (i.e., defined as between-groups mean squares) to within-groups variability (i.e., within-groups mean squares).

**TABLE 1.3** Single-Level Regression Analyses

|  | Reading | | Math | | Language | |
|---|---|---|---|---|---|---|
|  | Beta | SE | Beta | SE | Beta | SE |
| Female | –9.375 | 16.23 | –42.75 | 17.76 | –26.46 | 14.52 |
| SS Between | 301.339 (1 df) | | 6265.929 (1 df) | | 2400.149 (1 df) | |
| SS Within | 10836.375 (12 df) | | 12973.500 (12 df) | | 8672.208 (12 df) | |
| *F*-ratio | 0.334 ($p = 0.574$) | | 5.796 ($p = 0.033$) | | 3.321 ($p = 0.093$) | |
| *p* | 0.57 | | 0.03 | | 0.09 | |
| *R*-Square | 0.03 | | 0.33 | | 0.22 | |

To partition the variability in achievement, we disaggregate individuals' raw scores into their deviations about their respective group mean (within-groups variation) and disaggregate the group means from the overall grand mean (between-groups variation). This amounts to a key piece of information in a test of whether the difference in means in reading, for example, between males and females ($650.88 - 641.50 = 9.38$) in Table 1.2 is statistically significant in the population. As we noted, in Table 1.3 the regression coefficient representing the effect of gender on reading score ($\beta = -9.38, p > 0.10$) also summarizes the difference in means between males and females. As we would expect, the results using one-way ANOVA and multiple regression are consistent although presented in a slightly different fashion.

**Multivariate Analysis**

As suggested in Table 1.1, multivariate analysis is the more general case of univariate analysis; that is, it facilitates the examination of multiple independent and dependent variables in one simultaneous model. When we chose to examine the relationship between gender and each achievement test separately, our choice of analytic approach would have eliminated the possibility that students' reading scores were also correlated with their math and language scores. The initial correlations (not tabled), however, suggest that reading and math are highly correlated ($r = 0.79$), as are reading and language ($r = 0.76$) and language and math ($r = 0.85$).

*Multivariate Analysis of Variance*

We could use multivariate analysis of variance (MANOVA) to investigate whether gender affected student achievement more generally (i.e., using reading, math, and language test scores as dependent variables in the same model). The multivariate approach has the advantage of providing an analysis of differences considering all dependent variables simultaneously. It has the effect of controlling for the covariance structure between the dependent variables. Classic multivariate analysis uses descriptive information about means, standard deviations, and correlations (or covariances) to summarize relationships between the dependent variables in the sample initially. A set of means (called a vector, in matrix terminology) replaces the individual means for each achievement score in the model. The hypothesis tested is that the population from which the groups are selected has the same means for all dependent variables. A more sophisticated way to think about this system of three dependent variables is as representing a linear combination of the dependent variables, or a single latent (unobserved) variable that we might call
"achievement."

The MANOVA approach can be conceptualized as creating a latent (or underlying) achievement variable defined by a linear weighted function of the observed dependent variables and then assessing whether this function is the same for males and females. The scores on the dependent variables generated by the function can be seen as representing an individual's standing on a latent variable (Marcoulides & Hershberger, 1997). Although the achievement latent variable is corrected for correlations between the set of tests, from the MANOVA analysis we do not get any direct information about how strongly each test is associated with the underlying achievement factor. This information would help us understand how well each separate test contributes to the definition of the latent achievement variable. While we can obtain univariate information (e.g., parameter estimates, standard errors, hypothesis tests) about how predictors affect each dependent variable separately, this violates the multivariate nature of the outcome.

Matrices are important building blocks for both multivariate and multilevel analyses. A **covariance matrix** represents a convenient way to store information about observed variables (e.g., variances and covariances) that can be used to test relationships implied by a statistical model. Mathematical operations (e.g., multiplication or inversion) are performed on the covariance matrix as a way of determining whether a proposed set of relationships comprising a model explains patterns observed in the data. Remember that in the univariate case, the ratio of between-group variability to within-group variability is described by an *F*-ratio. This provides a test of the significance of difference between the means of two groups. In the multivariate case, there are sets of dependent variables, so a similar test of variability involves decomposing a total sample matrix of **sums of squares and cross products (SSCP)** into a between-subjects matrix and an error (or within-subjects) matrix. The question of whether or not there is significant variability in the outcomes due to groups is then answered with a multivariate test.

Summary measures of the variation that exists within a matrix are called **determinants**. The determinant represents a measure of generalized variance in the matrix after removing covariance. Ratios between determinants (similar to *F*-ratios) provide a test of the hypothesis about the effect of the independent variable on the linear combination of dependent variables. We can compare the within-groups portion of the variance to the total sample

SSCP matrix (between + within matrix) using a ratio of the respective determinants (|D|) of these matrices. One statistic often used for testing multivariate hypotheses is Wilks's lambda ($|D_W|/|D_T|$), which can be interpreted as a measure of the proportion of total variability in the outcomes not explained by group differences. It is of interest to note that the univariate case, for a single dependent variable, Wilks's lambda can

**TABLE 1.4** Multivariate Analysis of Variance (MANOVA) Results

|  | Within-Groups SSCP Matrix | | | Total SSCP Matrix | | | | | |
|---|---|---|---|---|---|---|---|---|---|
|  | Math | | Language | Math | | Reading | Language | | Reading |
| Reading | 10836.375 | 10214.750 | 7594.125 | 11137.714 | 11588.857 | 8444.571 | | | |
| Math | 10214.750 | 12973.500 | 8535.250 | 11588.857 | 19239.429 | 12413.286 | | | |
| Language | 7594.12 | 8535.250 | 8672.208 | 8444.571 | 12413.286 | 11072.357 | | | |
|  | Within determinant = $1.00888 \times 10^{11}$ | | | Total determinant = $2.27001 \times 10^{11}$ | | | | | |
|  | Wilks's $\Lambda = 1.00888 \times 10^{11} / 2.27001 \times 10^{11} = 0.444$ | | | | | | | | |

be expressed in terms of a ratio of the sum of squares within groups to the total sum of squares (Marcoulides & Hershberger, 1997).

The results of this analysis (Table 1.4) suggest that gender is significantly related to the latent achievement variable (Wilks's $\Lambda = 0.444$, $p = 0.037$). This result also suggests a conclusion somewhat inconsistent with the previous univariate regression analysis. Assuming that we felt the MANOVA approach was better suited to capture the theoretical relationships of interest in our population, we might then view the univariate regression results as incomplete and suggestive of an incorrect conclusion regarding the relationship between gender and academic achievement.

### *Structural Equation Modeling*

We could also conduct a multivariate analysis with structural equation modeling (SEM). SEM facilitates the specification and testing of models that include latent variables, multiple indicators, measurement errors, and complex structural relationships such as reciprocal causation. The SEM framework represents a generalization of both multiple regression and factor analysis and subsumes most linear modeling methods as special cases (Rigdon, 1998). SEM can be used to address two basic concerns in the example data: development of latent variables and the adjustments for measurement error in estimating these latent variables. As we shall next illustrate, SEM can be used to estimate well-known linear (e.g., ANOVA, MANOVA, multiple regression) models (Curran, 2003).

Defining constructs in terms of their observed indicators is generally the first part of an SEM analysis. This is often referred to as **confirmatory factor analysis (CFA)** since the proposed relationships are specified first and then examined against the data to see whether the hypothesized model is confirmed. This part of the analysis helps support the validity and reliability of proposed constructs through the measurement properties of their observed indicators. In the SEM approach to examining data structures, a smaller set of latent (unobserved) factors is hypothesized to be responsible for the specific pattern of variation and covariation present in a set of observed variables. In a technical sense, when a researcher tests a particular model, restrictions are imposed on the sample covariance matrix summarizing a set of empirical relationships (e.g., factor loadings, factor variances and covariances, residual covariances). Through such restrictions, the relationships observed in the sample data are compared to the restrictions defined through the mathematical model, and any of a number of assessments of "fit" between the data and the implied model can be derived. For example, a matrix of covariances among observed variables may be decomposed into a matrix of factor loadings and a matrix of errors, and the adequacy of the reproduced matrix of implied relationships may be examined against the data.

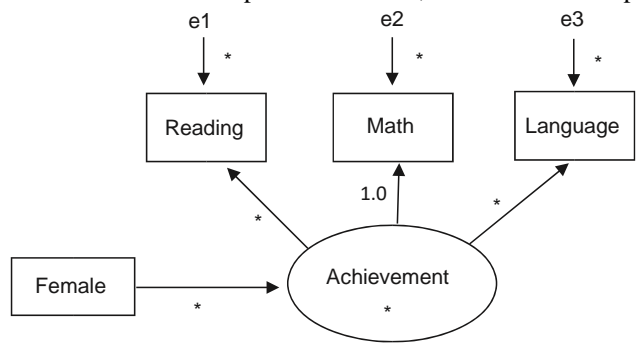

**FIGURE 1.2** Parameters estimated in proposed model of gender's influence on student achievement.

In the example summarized in Figure 1.2, we can treat the observed tests as if they define a latent achievement factor. Defining latent constructs through several observed indicators helps to address the second concern identified previously: producing more accurate estimates of structural relationships (i.e., regression coefficients) because the achievement factor has been corrected for measurement error. To provide a metric to measure the factor, we typically fix one factor loading to 1.0. In this case, we would also have to assume that the errors for each test do not covary (given our limited sample size). We can then examine whether the achievement results are different for males and females.

The structural model summarized in Figure 1.3 suggests that the three subtests are strong indictors of the latent achievement factor (i.e., with factor loadings ranging from 0.79 to 1.00). High factor loadings indicate that the achievement factor is well measured by the individual tests (i.e., corresponding errors for each indicator will be relatively small). Because we fixed one factor loading (Math) to 1.0, its corresponding standard error is not estimated (i.e., it is not tested for statistical significance). The unstandardized beta coefficient summarizing the effect of female on achievement is –42.73, and the standardized effect is –0.57 ($p < 0.05$). The remaining variance in student achievement unaccounted for by gender is summarized in parentheses (0.67), which implies that gender accounts for 33% of the variance in achievement in this small sample.

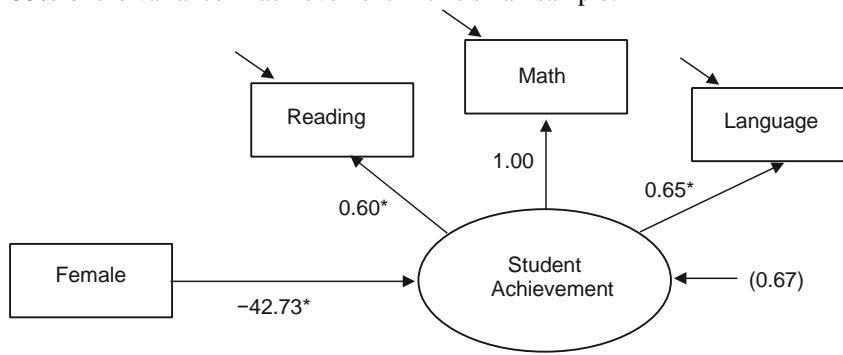

**FIGURE 1.3** Standardized and unstandardized (in parentheses) relationship between gender and student achievement (*$p < 0.05$).

## Summary

We have shown, through the example used in this chapter, the ways in which the multilevel model is simply an extension of ordinary single- level regression. In the last chapter, we called attention to the numerous shortcomings of ordinary regression when faced with a hierarchical data structure (e.g., unit of analysis questions and dependence of observations within units). This chapter has demonstrated some of the advantages of working in a multilevel framework rather than the traditional single- level approach. Beyond the basic model we provide here, the mixed- modeling routines permit a range of other specification and tests (including residual analyses) that can be used to illustrate some of the more complex conceptual elements we introduced in this chapter.

We have provided the basics of two- level multilevel regression modeling through the development of a series of models to explain variability in a random intercept and slope across level- 2 units. We illustrated model development by using the Mplus statistical software program. We paid considerable attention to the specification of Mplus input files and the interpretation of results. We also offered a fairly detailed consideration of how centering predictors can influence the interpretation of the model results.

Despite the clear advantages of multilevel mixed (or regression) models, limitations still remain. In the following chapters, we will extend the models we have worked with in Chapters 2 and 3 and consider structural equation modeling techniques that permit analysis of broader range of theoretical models and greater refinement of error specification. In the next chapter, we extend the basic two-level model to three levels and then also look at possible multivariate outcomes that can be specified within the mixed-model (or multilevel regression) framework. We cover some of the more complex specifications in subsequent chapters.

**Notes**

1. From a multilevel SEM perspective, the population covariance structure for individual $i$ in cluster $j$ would be

$$V(y_{ij}) = \Sigma_T = \Sigma_B + \Sigma_W. \quad (A3.1)$$

The within-group and between-group covariance structures can then be specified as the following:

$$\Sigma_W = \Lambda_W \Psi_W \Lambda'_W + \Theta_W$$
$$\Sigma_B = \Lambda_B \Psi_B \Lambda'_B + \Theta_B. \quad (A3.2)$$

The mean structure (specified at Level 2) is as follows:

$$\mu = \alpha_B + \Lambda_B \nu_B, \quad (A3.3)$$

where $\nu_B$ is a vector of latent factor means.

2. This level-1 model is often referred to as an unconditional model, that is, a level-1 model unconditional at Level 2.